\documentclass{aa}
\usepackage[varg]{txfonts}

\usepackage{natbib}
\bibpunct{(}{)}{;}{a}{}{,} 

\begin{document}

\title{Comparing population synthesis models of compact double white dwarfs to electromagnetic observations}

\author{Wouter G. J. van Zeist\inst{\ref{radb},\ref{leid},\ref{auck}}\thanks{Corresponding author: wouter.vanzeist@astro.ru.nl}
\and Jan van Roestel\inst{\ref{amst}}
\and Gijs Nelemans\inst{\ref{radb},\ref{sron},\ref{leuv}}
\and Jan J. Eldridge\inst{\ref{auck}}
\and Valeriya Korol\inst{\ref{mpa}}
\and Silvia Toonen\inst{\ref{amst}}
}

\institute{Department of Astrophysics/IMAPP, Radboud University, PO Box 9010, 6500 GL, Nijmegen, The Netherlands\label{radb}
\and Leiden Observatory, Leiden University, Einsteinweg 55, 2333 CC, Leiden, The Netherlands\label{leid}
\and Department of Physics, University of Auckland, Private Bag 92019, Auckland, New Zealand\label{auck}
\and Anton Pannekoek Institute for Astronomy, University of Amsterdam, 1090 GE, Amsterdam, The Netherlands\label{amst}
\and SRON, Netherlands Institute for Space Research, Niels Bohrweg 4, 2333 CA, Leiden, The Netherlands\label{sron}
\and Institute of Astronomy, KU Leuven, Celestijnenlaan 200D, B-3001, Leuven, Belgium\label{leuv}
\and Max Planck Institute for Astrophysics, Karl-Schwarzschild-Straße 1, 85748, Garching, Germany\label{mpa}
}

\date{Received XXX / Accepted YYY}

\abstract{Studies of the Galactic population of double white dwarfs (DWDs) that would be detectable in gravitational waves by LISA have found differences in the number of predicted detectable DWDs of more than an order of magnitude, depending on the  binary stellar evolution model used. Particularly, the binary population synthesis code \textsc{bpass} predicts 20 to 40 times fewer detectable DWDs than the codes \textsc{SeBa} or \textsc{bse}, which relates to differing treatments of mass transfer and common-envelope events (CEEs).}
{We aimed to investigate which of these models are closer to reality by comparing their predictions to the DWDs known from electromagnetic observations.}
{We compared the DWDs predicted by a \textsc{bpass} galaxy model and a \textsc{SeBa} galaxy model to a DWD catalogue and the sample of DWDs observed by the Zwicky Transient Facility (ZTF), taking into account the observational limits and biases of the ZTF survey.}
{We found that \textsc{bpass} underpredicts the number of short-period DWDs by at least an order of magnitude compared to the observations, while the \textsc{SeBa} galaxy model is consistent with the observations for DWDs more distant than 500 pc. These results highlight how LISA's observations of DWDs will provide invaluable information on aspects of stellar evolution such as mass transfer and CEEs, which will allow theoretical models to be better constrained.}
{}


\titlerunning{Comparing population synthesis models of compact DWDs to EM observations}
\authorrunning{W. G. J. van Zeist et al.}

\maketitle

\section{Introduction}

White dwarfs (WDs) are the final stage of evolution of relatively low-mass stars (lighter than $\sim$8 M$_{\odot}$). Close binaries consisting of two WDs (DWDs or WD–WD binaries) are of interest as they are expected to be sources of observable gravitational waves \citep[GWs;][]{lisa_astrophysics}.

While no DWDs have yet been observed in GWs, the LIGO–Virgo–KAGRA (LVK) Collaboration has detected numerous GW signals \citep[e.g.][]{gwtc3} from mergers of binaries containing black holes (BHs) and neutron stars (NSs), the remnants of stars with masses of at least 8$\sim$10 M$_{\odot}$, which are heavier than those that produce WDs. However, DWDs cannot be observed by LVK, as they have larger radii (on the order of 0.01 R$_{\odot}$), and consequently their mergers would occur at larger separations and thus lower frequencies (on the order of 0.1 Hz), which fall outside of the frequency band of the LVK detectors. Instead, DWDs could be detected by a GW observatory operating in a lower frequency band, such as the in-development Laser Interferometer Space Antenna \citep[LISA;][]{lisa_l3,lisa_redbook}.

Unlike the ground-based LVK detectors, LISA will be a space-based observatory, consisting of three satellites in a triangular formation with a separation, and therefore interferometer arm length, of 2.5 million km \citep[see][]{lisa_redbook}. Because of this arm length, LISA will be sensitive to GWs at frequencies lower than those of LVK, between about 0.1 mHz and 1 Hz \citep{lisa_l3,lisa_redbook}. DWDs emit GWs in this range during their inspiral stage prior to merging, making Galactic DWDs observable targets for LISA \citep[e.g.][]{hils1990,schneider2001,nelemans2001,ruiter2010,nissanke2012,korol2017,lamberts2019,breivik2020,li2020,li2023,bpassmilkyway}; overviews of studies on the potential detection of DWDs and other compact binaries by LISA are given in Sect. 1 of \citet{lisa_astrophysics} and Sect. 4.3 of \citet{breivik2025}.

The aforementioned studies about the Galactic population of DWDs that LISA may detect generally make use of binary stellar evolution models. These are combined with models of the star formation history (SFH) and structure of the Milky Way (MW) to create a synthetic population of DWDs embedded in a model galaxy. As LISA is not operational yet, there are no GW observations to compare these models to, but there have been detections of DWDs in electromagnetic (EM) observations \citep[see e.g.][]{kupfer2024}. However, the EM-observed sample of DWDs is limited to the part of the MW closest to Earth due the intrinsic faintness of WDs, and is affected by numerous observational biases which affect the chances of detecting the binarity of the systems and even seeing the light of the WDs in the first place. We note that \citet{korol2022} created a model of the DWD throughout the MW based just on EM observations as opposed to population synthesis, using the methods of \citet{maoz2012,maoz2018}. However, this model depends on several restrictive assumptions about the period distribution and particularly the WD mass distribution. These assumptions lead to the \citet{korol2022} DWD population having quite high chirp masses (averaging around 0.5 M$_{\odot}$) compared to models like \citet{korol2017} and \citet{lamberts2019}, which have DWD chirp masses on average around 0.3 M$_{\odot}$; these higher chirp masses would lead to the GW signals of these binaries being comparatively louder, as the strain amplitude of a binary's GW signal scales with its chirp mass to the 5/3rd power \citep[see e.g.][]{krolak2004}.

The specific biases affecting an EM detection depend on the detection method used. For example, one common way of detecting the binarity of a system is by finding dips in its lightcurve corresponding to the objects in the binary eclipsing each other. This requires the system to have an inclination angle within a certain range such that the objects pass in front of each other from the perspective of the observer, and the width of this range depends on the objects' radii and separation \citep[see e.g.][]{first_eclipsing_dwd}. However, DWD binaries can also be detected by measuring the Doppler shift in the WD's spectra due to the binary's radial velocity variations, which does not require an eclipse, but the Doppler shift does depend on the inclination and the orbital velocity \citep[e.g.][]{robinson1987,marsh1995,napiwotzki2001,elm_survey_2,elm_survey_8,munday2024}.

There are also additional observational biases that affect the detection of DWDs in EM. The apparent brightness of the DWD must be high enough for it to be visible to the telescope, so more distant or more dust-obscured DWDs are more difficult to detect. Additionally, the apparent brightness depends on the intrinsic luminosity of the WDs, which itself depends on their masses, radii and ages, as WDs become fainter as they age. However, recent WD-focused EM surveys such as the ELM Survey \citep{elm_survey_1,elm_survey_9} and the Zwicky Transient Facility \citep[ZTF;][]{ztf_overview} have provided larger, more homogeneous samples of DWDs than were previously known, meaning a comparison between theoretical models and observations is now possible.

Predictions of the DWD population detectable by LISA vary significantly between population synthesis codes, which each use different methods and assumptions to model stellar evolution. In particular, \citet{wouter_clusters_clouds} and \citet{bpassmilkyway} found that the stellar evolution/population synthesis code \textsc{bpass} \citep{bpass1,bpass2} predicts a few tens of times fewer detectable DWDs than other codes like \textsc{bse} \citep{hurley2002} and \textsc{SeBa} \citep{portegies1996,nelemans_seba,seba_toonen}.

Specifically, \citet{bpassmilkyway} compared a \textsc{bpass} MW model to the \textsc{bse} MW model of \citet{lamberts2019}, and \citet{wouter_clusters_clouds} compared \textsc{bpass} predictions for the Magellanic Clouds to the \textsc{SeBa} predictions of \citet{lmclisa}, with both studies finding a few tens of times fewer LISA-detectable DWDs in the \textsc{bpass} models compared to \textsc{bse} and \textsc{SeBa}. However, \citet{bpassmilkyway} did find that the space density of DWDs predicted by the \textsc{bpass} galaxy model is consistent with the sample of observed WDs within 25 pc from \citet{holberg2016}, which \textsc{SeBa} is also consistent with \citep{toonen2017}.

In the \citet{bpassmilkyway} MW model, there were 600–700 LISA-detectable DWDs in total, compared to 13,000–14,000 in the \citet{lamberts2019} MW model when evaluated with the same detectability criteria. By comparison, \citet{ruiter2010}, using the population synthesis code \textsc{StarTrack} \citep{startrack_1,startrack_2}, predicted 11,000 resolved LISA DWDs in the MW (though they used an earlier LISA design with longer arms and greater sensitivity than the current design, so this number would be lower with the current design); \citet{korol2017}, using \textsc{SeBa}, predicted 25,000; and \citet{delfavero2025}, using \textsc{cosmic} \citep{cosmic}, predicted on the order of 10,000. Notably, \textsc{SeBa} predicts about twice as many LISA-detectable DWDs as \textsc{bse}, \textsc{StarTrack} and \textsc{cosmic}.

\citet{wouter_clusters_clouds} found that the lower number of resolved DWDs in \textsc{bpass} compared to \textsc{SeBa} and \textsc{bse} is not due to the total numbers of DWDs predicted being fewer, but the result of differences in the period distribution of the DWDs output by the different codes, with \textsc{bpass} producing fewer high-frequency binaries than \textsc{bse} and \textsc{SeBa}. This discrepancy in turn results from different treatments of mass transfer stability and common-envelope events (CEE), which cause DWDs in \textsc{bpass} to be wider at the end of a CEE than in \textsc{bse} and \textsc{SeBa}.

In this study, we took two models of the DWD population of the MW, a \textsc{bpass}-based model from \citet{bpassmilkyway} and a \textsc{SeBa}-based model from \citet{korol2017}, and compared these to catalogues of EM observations of DWDs, in order to evaluate how realistic these models are. Particularly, we looked at the period distribution as this is the key factor in the diverging LISA predictions between \textsc{bpass} and \textsc{SeBa} as reported in \citet{bpassmilkyway} and \citet{wouter_clusters_clouds}. While we did not perform an exhaustive modelling of all the potential selection effects that could be affecting the observational samples, we aimed to investigate upper limits on the numbers of observable systems in the model populations and compare these to the number of real observations. We note that there have been previous studies that have compared population synthesis models to EM observations of DWDs \citep[e.g.][]{nelemans_gamma1,toonen2017,li2023}, but these have focused primarily on wide binaries that would not be detectable by LISA, while we focus on shorter-period systems.

This paper is structured as follows: in Sect. \ref{method_empaper} we explain our methods of comparing population synthesis models to EM observations of DWDs, in Sect. \ref{results_empaper} we present our results, in Sect. \ref{discussion_empaper} we discuss the uncertainties and limitations of these results and in Sect. \ref{conclusions_empaper} we state our conclusions.

\section{Method} \label{method_empaper}

\subsection{Population synthesis models}

The two population synthesis models we compared each consist of synthetic DWDs that are spatially distributed across a model of the MW.

The \textsc{bpass} galaxy model is taken from \citet{bpassmilkyway}. The SFH and structure of the MW model are taken from the galaxy formation model \textsc{fire} \citep{hopkins_fire}, specifically the ``m12i'' simulation from \citet{wetzel2023}. The specific version of \textsc{bpass} used in the stellar population synthesis (v2.2.1) is described in Sect. 2.2 of \citet{bpassmilkyway} and Sect. 2.3 of \citet{wouter_gw_spectral}. The population uses an initial mass function from \citet{kroupa1993} and initial binary parameters from \citet{moe2017}. The GW sources are output by the \textsc{bpass} module \textsc{tui}, which is described in \citet{bpassmassdist}, \citet{bpassgw170817}, \citet{bpassmasstransfer} and \citet{wouter_gw_spectral}.

The \textsc{SeBa} galaxy model is taken from \citet{korol2017}. The SFH and structure of the MW model are taken from a code developed by \citet{nelemans_seba,nelemans2004} and \citet{toonen2013} based upon the galaxy evolution model of \citet{boissier1999}. \citet{korol2017} produced two versions of their galaxy model with different CEE assumptions, named ``$\alpha\alpha$'' and ``$\gamma\alpha$'' respectively. The ``$\gamma\alpha$'' model typically uses the $\gamma$-formalism \citep{nelemans_gamma1,nelemans_gamma2} for CEEs during the first binary interaction and the $\alpha$-formalism \citep{webbink1984} for the second, while the ``$\alpha\alpha$'' model uses the $\alpha$-formalism in both cases; we discuss CEE prescriptions in more detail in Sect. \ref{discussion_cee}. We used the ``$\gamma\alpha$'' version of the galaxy model. The specific version of \textsc{SeBa} used is described in \citet{seba_toonen} and \citet{toonen2013}. The population uses an initial mass function from \citet{kroupa1993} and initial binary parameters from \citet{heggie1975} and \citet{abt1983}; these initial conditions are not identical, but quite similar to those of the \textsc{bpass} galaxy model.

\subsection{EM observations of DWDs} \label{method_em_obs}

\begin{figure}
    \centering
    \includegraphics[width=\columnwidth]{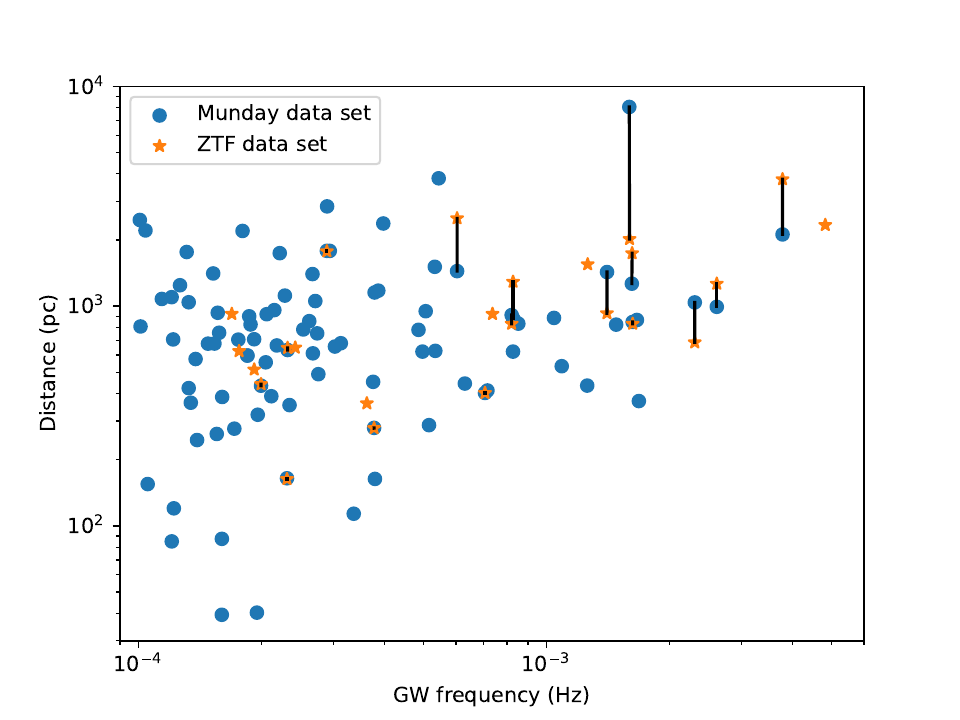}
    \caption{GW frequency and distance of the binaries in the Munday and ZTF datasets. When a binary appears in both datasets, the two corresponding points are connected with a line; the distances differ due to the use of different distance measurement techniques.}
    \label{em_data_set_plot}
\end{figure}

\begin{table*}
    \centering
    \caption{Binary parameters of the DWDs in the ZTF dataset.}
    \begin{tabular}{c c c c c}
    Name & P$_{\rm orb}$ (min) & log(f$_{\rm GW}$/Hz) & D (pc) & References \\
    \hline
    ZTF J1539+5027 & 6.91 & -2.317 & 2340 & \citet{burdge2019} \\
    ZTF J2243+5242 & 8.80 & -2.422 & 3778 & \citet{ztf_8min} \\
    SDSS J0651+2844 & 12.75 & -2.583 & 1265 & \citet{brown2011} \\
    ZTF J0538+1953 & 14.44 & -2.637 & 684 & \citet{ztf_wdwd_cat} \\
    ZTF J0526+5934 & 20.51 & -2.789 & 831 & \citet{ztf_wdwd_cat,kosakowski2023} \\
    PTF J0533+0209 & 20.57 & -2.790 & 1740 & \citet{ztf_wdwd_cat} \\
    ZTF J2029+1534 & 20.87 & -2.797 & 2020 & \citet{ztf_wdwd_cat} \\
    ZTF J0722–1839 & 23.71 & -2.852 & 928 & \citet{ztf_wdwd_cat} \\
    ZTF J1749+0924 & 26.43 & -2.899 & 1550 & \citet{ztf_wdwd_cat} \\
    SDSS J0822+3048 & 40.28 & -3.082 & 1291 & \citet{brown2017} \\
    ZTF J1901+5309 & 40.60 & -3.086 & 831 & \citet{coughlin2020} \\
    ZTF J0720+6439 & 45.23 & -3.133 & 921 & \citet{vanroestel2025} \\
    WD J0225–6920 & 47.19 & -3.151 & 403 & \citet{munday2023} \\
    ZTF J2320+3750 & 55.25 & -3.219 & 2510 & \citet{ztf_wdwd_cat} \\
    ZTF J0221+1710 & 88.25 & -3.423 & 271 & \citet{elm_survey_south_2,vanroestel2025} \\
    ZTF J1356+5705 & 91.93 & -3.441 & 361 & \citet{vanroestel2025} \\
    SDSS J0751–0141 & 115.22 & -3.539 & 1776 & \citet{kilic2014} \\
    ZTF J2249+0117 & 137.75 & -3.616 & 647 & \citet{vanroestel2025} \\
    SDSS J1152+0248 & 143.80 & -3.635 & 645 & \citet{hallakoun2016} \\
    J2102–4145 & 144.30 & -3.636 & 163 & \citet{elm_survey_south_2,antunes2024} \\
    CSS 41177 & 167.06 & -3.700 & 441 & \citet{parsons2011,bours2014} \\
    ZTF J1110+7445 & 173.63 & -3.717 & 514 & \citet{vanroestel2025} \\
    ZTF J1758+7642 & 189.12 & -3.754 & 624 & \citet{vanroestel2025} \\
    ZTF J0238+0933 & 197.03 & -3.772 & 921 & \citet{vanroestel2025} \\

    \end{tabular}
    \tablefoot{Binary parameters of the 24 DWDs in our ZTF dataset. P$_{\rm orb}$ is the orbital period, f$_{\rm GW}$ is the GW frequency and D is the distance. Spectroscopic distances from ZTF observations \citep{burdge2019,ztf_wdwd_cat,coughlin2020} are listed where available, and otherwise \textit{Gaia} parallax-derived distances \citep{bailerjones2021} are given.}
    \label{ztf_table}
\end{table*}

We compared the DWDs in the population synthesis MW models to two catalogues of EM observations of DWDs, which we will refer to as the ZTF dataset and the Munday dataset, respectively.

The ZTF dataset consists of a set of 24 DWDs that have been observed by ZTF, an optical time-domain survey using the Palomar telescope \citep{ztf_overview}. These binaries have been detected predominantly through eclipses. The binary parameters of these systems have previously been published \citep{brown2011,parsons2011,bours2014,kilic2014,hallakoun2016,brown2017,burdge2019,ztf_8min,ztf_wdwd_cat,coughlin2020,elm_survey_south_2,kosakowski2023,munday2023,antunes2024,vanroestel2025}, but as there is no single reference collating these, we summarise their binary parameters in Table \ref{ztf_table}.

The Munday dataset is a subset of a catalogue from \citet{munday2024}, which aims to be a comprehensive listing of all EM-observed DWDs in the literature, including the observations from the ZTF surveys but also from numerous other surveys and telescopes, such as the ELM Survey \citep{elm_survey_1,elm_survey_9} and the SPY survey \citep{spy_survey}. It also encompasses the data of earlier DWD catalogues such as those of \citet{kruckow2021} and \citet{40pc_wdwd}.

In our Munday dataset, we use those binaries from the \citet{munday2024} catalogue (specifically, the version as of 21 November 2024) for which both a period and a distance are given and that have a GW frequency of at least 0.1 mHz. From the remaining sample, we filter out a few specific systems: ZTF J0546+3843 and ZTF J1858–2024, as these are mass-transferring DWDs \citep{chakraborty2024}; and ZTF J1946+3203 \citep{ztf_wdwd_cat} and J2049+3351 \citep{elm_survey_south_2}, as there is a chance that these may not actually be DWDs, but rather systems consisting of a WD and a subdwarf. Additionally, while ZTF J1758+7642 was confirmed as a DWD by \citet{vanroestel2025}, in the \citet{munday2024} catalogue it is listed as a DWD with an unknown companion \citep[based on the data of][]{keller2022}, so we do not include it in our Munday dataset. For the system J0526+5934, it has also been suggested that it might be a binary consisting of a WD and a hot subdwarf \citep{kosakowski2023,lin2024}, but \citet{rebassa2024} argue that it is most likely to be a DWD, so we do include it in our dataset. With these restrictions, we obtain a final dataset of 94 DWDs.

Of the 24 DWDs in the ZTF dataset, 16 are also in the Munday dataset. Of the remaining eight systems, six are from \citet{vanroestel2025} (including the aforementioned ZTF J1758+7642), which was published more recently than the version of the \citet{munday2024} catalogue we used; and the final two systems are ZTF J1539+5027 and ZTF J1749+0924, which are listed without a distance measurement in the Munday catalogue and so are not included in our filtered subset. The frequencies given are the same for both datasets, but the distances given for each binary differ by up to a factor of a few due to the usage of different methods of distance measurement: in the Munday dataset, all distances are derived from \textit{Gaia} parallaxes \citep{gaia_dr3}, but in the ZTF dataset, spectroscopic distances from ZTF observations \citep{burdge2019,ztf_wdwd_cat,coughlin2020} are used where available, and parallax-derived distances otherwise; the reason for preferring the spectroscopic distances is that, due to the faintness of the WDs in the sample, the \textit{Gaia} parallaxes are rather uncertain, making the parallax-derived distances less reliable. However, we note that the parallax-derived distances will improve with future \textit{Gaia} releases.

Fig. \ref{em_data_set_plot} shows the set of binaries that we used from both datasets, in terms of GW frequency ($f_{\rm GW} = 2/P_{\rm orb}$) and distance. It can be seen that the binaries range in distance from 40 to 10,000 pc, and in frequency from 0.1 to 5 mHz. The \citet{munday2024} catalogue extends to lower frequencies, but the binaries below 0.1 mHz will not be detectable by LISA, so we do not include them in our sample. The black lines show the binaries that are in both datasets, but with differing distance measurements.

\subsection{Correcting for observational biases} \label{method_biases}

When comparing the galaxy models to the observed DWDs, we needed to take into account that the observed samples do not contain all Galactic DWDs, but only those that are both bright enough to be detected by EM telescopes and identifiable as binaries via eclipses or radial velocity measurements. Therefore, to get an indication of whether a model is overpredicting or underpredicting compared to reality, we needed to calculate how many of the DWDs predicted by the model would actually be detectable, for which an understanding of the observational biases is required.

A detailed modelling of all the relevant observational selection effects is highly complicated. However, given the very large discrepancy between the population synthesis models, we aimed to investigate some upper limits on the number of observable systems in the models and compare those to the real observed number.

Here we note some differences between our two observational datasets: the ZTF dataset mostly consists of DWDs detected through eclipses, while the Munday dataset includes both DWDs detected through eclipses and ones detected through radial velocity measurements. Additionally, as stated in Sect. \ref{method_em_obs}, the ZTF dataset is based on observations from a single telescope that has covered a significant fraction of the sky in a homogeneous way, while the Munday dataset uses multiple telescopes with inhomogeneous sky and parameter coverage. For these reasons, the observational biases affecting each dataset are different, and those of the ZTF dataset are simpler to calculate numerically, and so for the remainder of this section as well as for the corresponding results in Sect. \ref{results_biases}, we only used the ZTF dataset.

\subsubsection{Probability of eclipsing} \label{method_eclipse}

\begin{figure}
    \centering
    \includegraphics[width=\columnwidth]{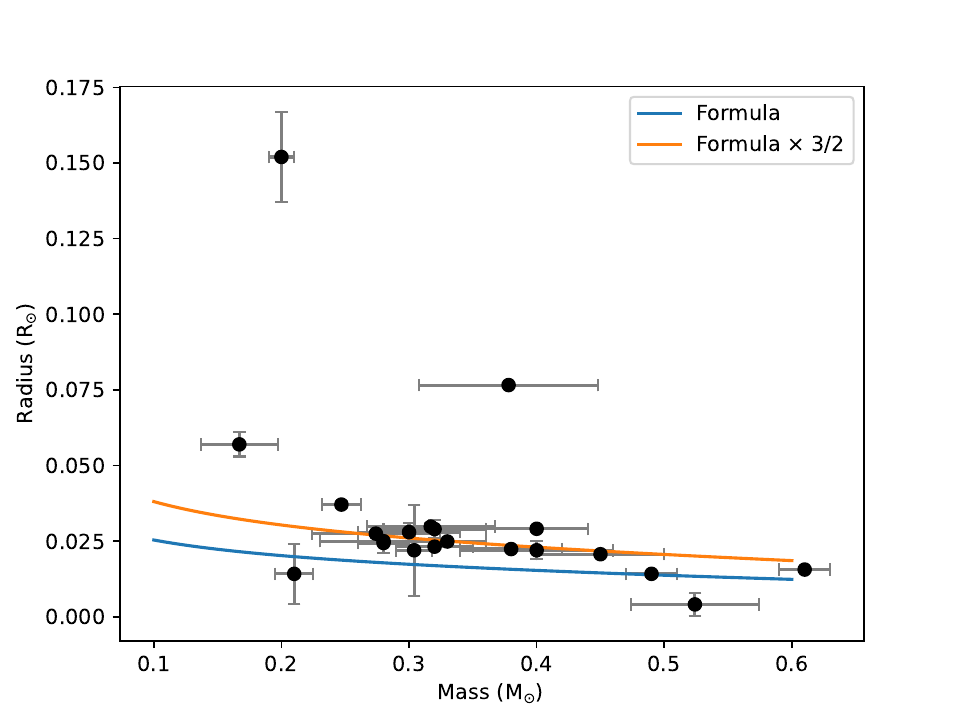}
    \caption{Comparison of the WD radii predicted by Eq. \ref{radius_eq} to the WDs in the ZTF DWD dataset for which both a mass and a radius are listed. The black points are the observed WDs, with measurement uncertainties. The blue line is the radii as predicted by Eq. \ref{radius_eq}, and the orange line is those radii multiplied by 3/2. The two outliers with the highest radii might be extremely low-mass WDs \citep{ztf_wdwd_cat}, or could be hot subdwarfs instead of WDs \citep{kosakowski2023,lin2024,rebassa2024}.}
    \label{mass_radius_plot}
\end{figure}

The DWDs in the ZTF dataset have been identified as binaries through eclipses; that is, one of the WDs passing in front of the other from the perspective of the Earth, occluding its light and causing a dip in the lightcurve. There are two exceptions in the dataset, PTF J0533+0209 and ZTF J2320+3750 \citep{ztf_wdwd_cat}, which are not eclipsing but were identified as binaries through ellipsoidal variability in their lightcurves \citep[see e.g.][]{morris1985}, though their inclinations are similar to those of the eclipsing systems in the sample of \citet{ztf_wdwd_cat}.

Eclipses are only visible if the plane of the binary's orbit is oriented in such a way that the WDs pass in front of each other from the perspective of the Earth. The range of orientations for which an eclipse occurs, and therefore the probability that a randomly oriented binary will be eclipsing, is proportional to $P^{-2/3}$, where $P$ is the orbital period of the binary \citep[e.g.][]{vanroestel2018,elbadry2022}; that is, short-period binaries are more likely to be eclipsing than longer-period binaries, introducing a period/frequency bias into the observations.

Specifically, from geometrical considerations one finds that a binary will be eclipsing from the perspective of the Earth if:
\begin{equation} \label{eclipse_eq}
    \cos \iota < \frac{R_1 + R_2}{a},
\end{equation}
where $\iota$ is the inclination of the binary's orbital plane with respect to the Earth, $a$ is the binary's semi-major axis and $R_1$ and $R_2$ are the radii of the two binary components, respectively. Assuming a binary is oriented randomly, then $(R_1 + R_2)/a$ gives the probability of that binary being an eclipsing one.

The relation between a WD's mass, which we know for the population synthesis models, and its radius is quite complex and non-linear. One commonly-used formulation is given by \citet{verbunt1988} based upon calculations by Peter Eggleton:
\begin{equation} \label{radius_eq}
    \begin{split}
    R = 0.0114 \left[\left(\frac{M}{M_{\rm Ch}}\right)^{-2/3} - \left(\frac{M}{M_{\rm Ch}}\right)^{2/3}\right]^{1/2} \\
    \times \left[1 + 3.5\left(\frac{M}{M_p}\right)^{-2/3} + \left(\frac{M}{M_p}\right)^{-1}\right]^{-2/3}.
    \end{split}
\end{equation}
In this equation, $R$ is the WD radius in R$_{\odot}$, $M$ is the WD mass in M$_{\odot}$, $M_{\rm Ch}$ is the Chandrasekhar mass (1.44 M$_{\odot}$) and $M_p$ is a constant with a value of $5.7 \times 10^{-4}$ M$_{\odot}$.

We note that \citet{parsons2017} found that the (zero-temperature) \citet{verbunt1988} mass-radius relation may underpredict WD radii, particularly for WDs with low masses and/or high temperatures. Similarly, when we compared the radii produced by Eq. \ref{radius_eq} to the WD masses and radii actually measured in the ZTF dataset, we found that the radii predicted by Eq. \ref{radius_eq} were generally smaller than those actually measured; this is illustrated in Fig. \ref{mass_radius_plot}, where the points are those WDs from the ZTF dataset for which both a mass and a radius are listed, and the blue line shows the radii predicted by Eq. \ref{radius_eq} for WD masses between 0.1 and 0.6 M$_{\odot}$. We also plot, in the orange line, the radii from Eq. \ref{radius_eq} multiplied by a factor of 3/2; by manual inspection, this line fits most of the ZTF WDs reasonably well, apart from a few outliers that we discuss in the next paragraph. Due to scatter and uncertainty in the ZTF mass-radius measurements, the ``true'' coefficient could potentially be somewhat larger or smaller, though probably not by more than a factor of two, which is small compared to the order-of-magnitude differences between the population models. Hence, we judged that additional statistical analysis would not have significant benefit, and used a value of 3/2 for this coefficient. Therefore, we calculated the probability of a binary in the model populations being an eclipsing one with Eq. \ref{eclipse_eq}, using the radii from Eq. \ref{radius_eq} multiplied by 3/2. We use this eclipse probability to weight each object when we sum the total number of DWDs in a frequency-distance bin in Sect. \ref{results_biases}.

There are some outliers in Fig. \ref{mass_radius_plot} with larger radii than predicted by our fit to the mass-radius relationship. However, we note that the two WDs with the largest radii in Fig. \ref{mass_radius_plot} are unusual objects. The largest WD belongs to the binary ZTF J2320+3750 and may be an extremely low-mass WD \citep{ztf_wdwd_cat}. The second-largest WD belongs to ZTF J0526+5934; as mentioned previously, it is not certain whether this object is actually a WD or a hot subdwarf \citep{kosakowski2023,lin2024}, though \citet{rebassa2024} argue that it is most likely to be a WD, just with a lower mass than in the ZTF data.

\subsubsection{ZTF sky coverage} \label{method_coverage}

The ZTF survey is based upon the data from a single telescope facility, and so it does not cover the full sky. This reduces the probability of a given DWD being detected: even if it is eclipsing, if it is located in a part of the sky ZTF cannot see, then it will not be detected.

To estimate the probability that a given DWD is within ZTF's sky coverage, we first noted that 75\% of the sky is visible from the Palomar telescope facility used by ZTF. The primary grid of ZTF pointings, used for ZTF's DWD survey, covers approx. 88\% of the sky visible from ZTF \citep{ztf_overview,ztf_surveys}. This grid is divided into 636 fields, of which 598 have been observed at least 600 times, which is sufficient to be considered well-sampled for the purposes of detecting eclipsing DWDs \citep{vanroestel2022}. Thus, the fraction of the sky that is well-sampled by ZTF is $0.75 \times 0.88 \times \frac{598}{636} \approx 0.62$.

The preceding calculation ignores that DWDs are not distributed isotropically through the sky, but are more concentrated in the Galactic disk. An alternative way of formulating the sky coverage is to take each of the WDs in the \textit{Gaia} all-sky catalogue \citep{gaia_wdcat} and find how many times their sky location has been observed by ZTF. Then, the fraction of WDs that have been observed at least 600 times is approx. 0.54.

The difference between these two values is small compared to the orders-of-magnitude differences between the population synthesis models, so for simplicity we used a value of 0.6 for the ZTF sky coverage. We used this value for the detectability-corrected population synthesis predictions in Sect. \ref{results_biases}: once we calculated the number of DWDs in the population synthesis models that would be eclipsing, we multiplied the total in each frequency bin by 0.6 to reflect the number of these that would be located within ZTF's sky coverage.

\subsubsection{ZTF magnitude limit} \label{method_mag_limit}

Aside from the eclipse probability and ZTF's sky coverage, another factor that limits how many of the DWDs from the population synthesis models would actually be detectable is brightness. That is, ZTF can only detect binaries if they are sufficiently bright -- specifically, if they are brighter than the limiting magnitude of the detector. For ZTF, the limiting magnitude is 21.1--20.8 in \textit{g}-band and 20.9--20.6 in \textit{r}-band \citep{ztf_overview}; however, identifying the binarity of a DWD is more difficult than simply detecting light, and all of the DWDs in the ZTF dataset have a magnitude of 20.5 or brighter.

Taking into account that not all DWDs in the model population would be bright enough to be detectable would further lower the predicted numbers of detectable DWDs in the models. We investigated this by taking the \textsc{SeBa} galaxy model of \citet{korol2017} and computed a new model population by only including those DWDs brighter than a certain threshold (for which we took an \textit{r}-band magnitude of 20.5 as a default value). 

The WD magnitudes were calculated within \textsc{SeBa} as described in \citet{toonen2013}, using WD cooling models from \citet{holberg2006,kowalski2006,tremblay2011}. We computed the WD absolute magnitudes as a function of WD cooling time and WD mass via interpolation in the cooling model tables. In this procedure, the WD radius is not explicitly used, so the adaptation of the WD radii that we use to calculate the eclipse probabilities does not affect the calculation of the magnitude. The question is if the increase in radius we find observationally would have an influence on the cooling models, but this is beyond the scope of this paper. The effect on the magnitude as function of cooling age is also not simple: a larger radius leads to larger luminosity for given effective temperature, but also to faster cooling, i.e. lower effective temperature and luminosity at a given age.

Galactic extinction was applied to the WD magnitudes using the model of \citet{sandage1972} as implemented in \citet{nelemans2004}. We added together the \textit{r}-band magnitudes of the two components as follows, to compute a total magnitude which we compared to the threshold:
\begin{equation}
    m_{\rm tot} = -2.5 \log \left(10^{\frac{m_1}{-2.5}} + 10^{\frac{m_2}{-2.5}}\right).
\end{equation}
In this equation, $m_1$ and $m_2$ are the magnitudes of the two components and $m_{\rm tot}$ is the combined magnitude.

\section{Results} \label{results_empaper}

\subsection{DWDs by frequency and distance}

\begin{figure*}
    \centering
    \includegraphics[width=\columnwidth]{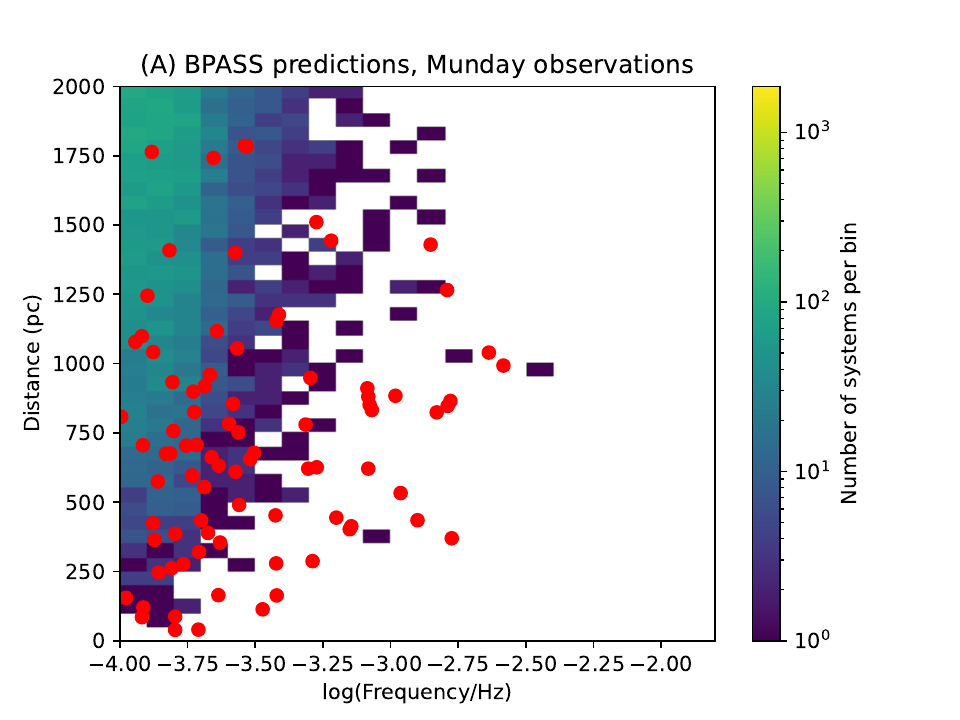}
    \includegraphics[width=\columnwidth]{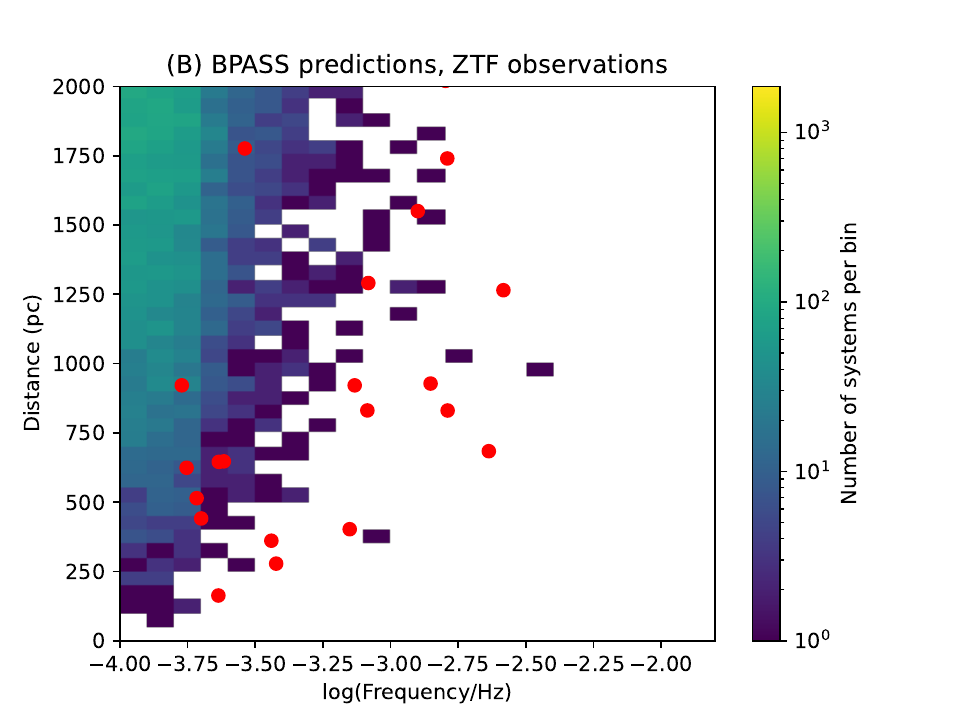}
    \includegraphics[width=\columnwidth]{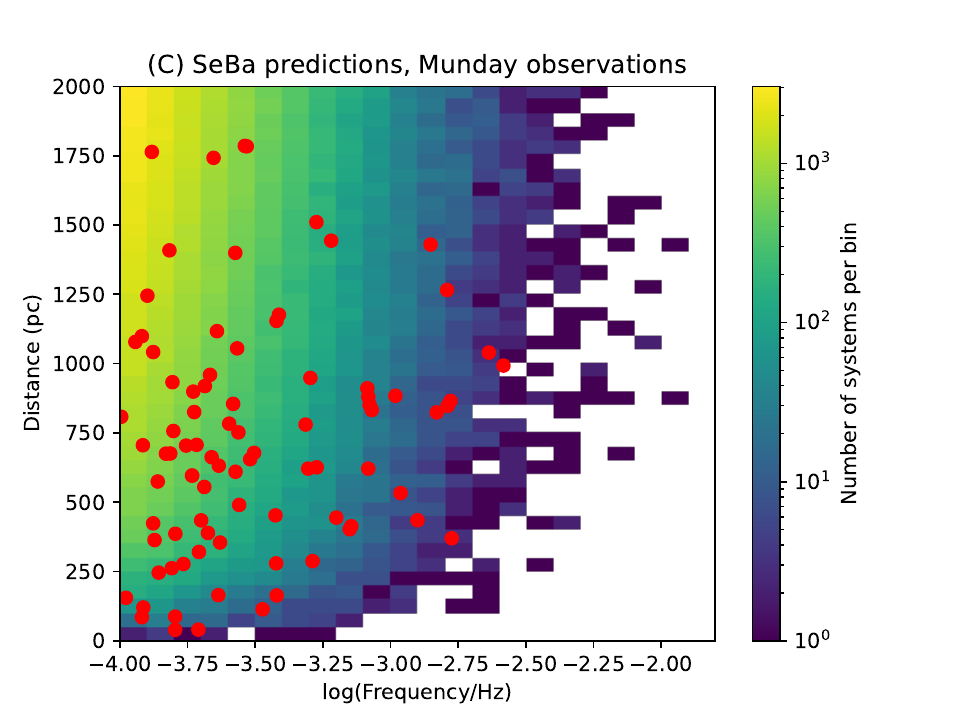}
    \includegraphics[width=\columnwidth]{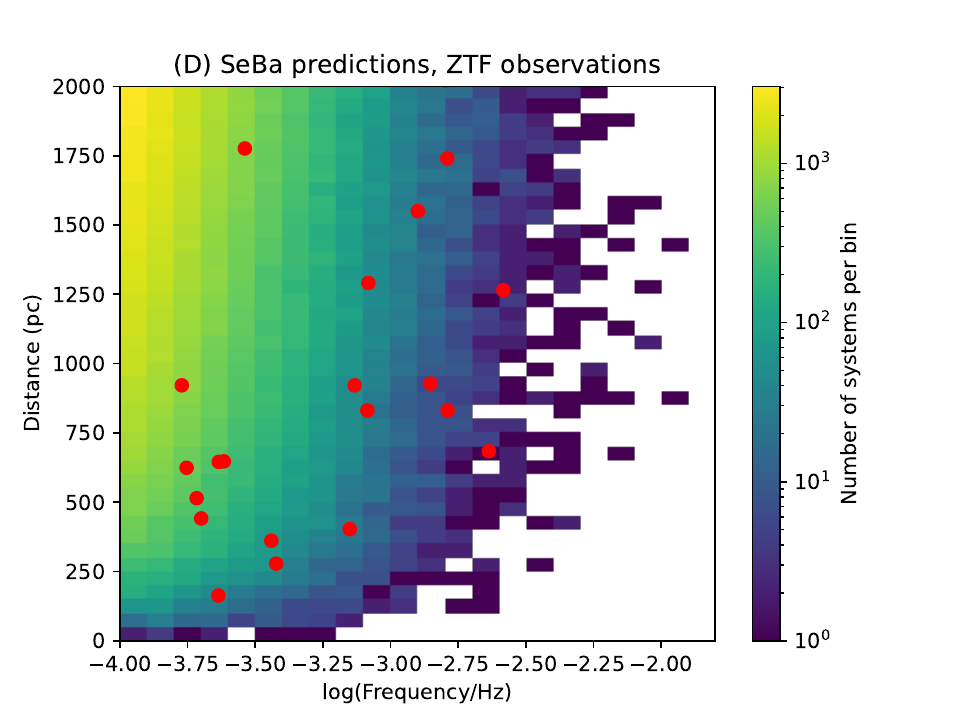}
    \caption{Comparison of the \textsc{bpass} (panels A and B) and \textsc{SeBa} (panels C and D) model populations to the observed systems from the Munday dataset (panels A and C) and the ZTF dataset (panels B and D), in terms of the GW frequency and distance of the binaries. Each rectangular cell represents a frequency/distance bin and is shaded according to the number of systems the population synthesis model predicts for that bin; white means zero systems are predicted. Each red dot is an observed system.}
    \label{bpass_seba_freqdist}
\end{figure*}

The top two panels of Fig. \ref{bpass_seba_freqdist} plot the DWDs present in the \textsc{bpass} galaxy model in terms of the frequency and distance of the systems, compared to the observed systems from the Munday and ZTF datasets. The bottom panels show the same, but using the \textsc{SeBa} galaxy model.

By comparing these figures, it is immediately apparent that the \textsc{SeBa} galaxy model contains far more high-frequency DWDs than the \textsc{bpass} galaxy model, affirming the conclusions of \citet{bpassmilkyway} and \citet{wouter_clusters_clouds}. For both galaxy models, the total number of DWDs in the model is orders of magnitude higher than the number of observed DWDs, but this is expected due to the limitations of the observations.

However, for the \textsc{bpass} galaxy model, we can see in the top row of Fig. \ref{bpass_seba_freqdist} that at the highest frequencies (> $10^{-3}$ Hz), the total number of predicted binaries is of comparable number to the observed binaries. However, as in this comparison we are not taking into account the selection effects that limit the EM observations, we would expect the number of predicted binaries to be significantly higher than the number of observed binaries (except possibly at the closest distances), as we see for the \textsc{SeBa} galaxy model in the bottom row of Fig. \ref{bpass_seba_freqdist}. This suggests the \textsc{bpass} galaxy model may be underpredicting the number of high-frequency DWDs; we investigate this in more detail in the next sections.

\subsection{Model predictions of DWDs adjusted for observational biases} \label{results_biases}

\begin{figure*}
    \centering
    \includegraphics[width=0.33\linewidth]{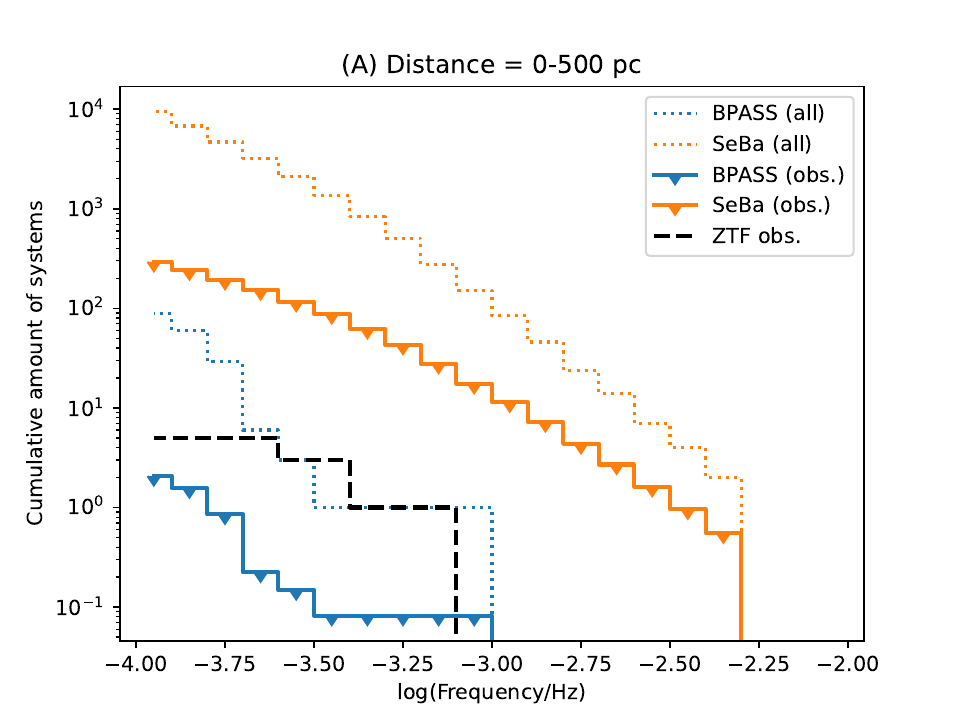}
    \includegraphics[width=0.33\linewidth]{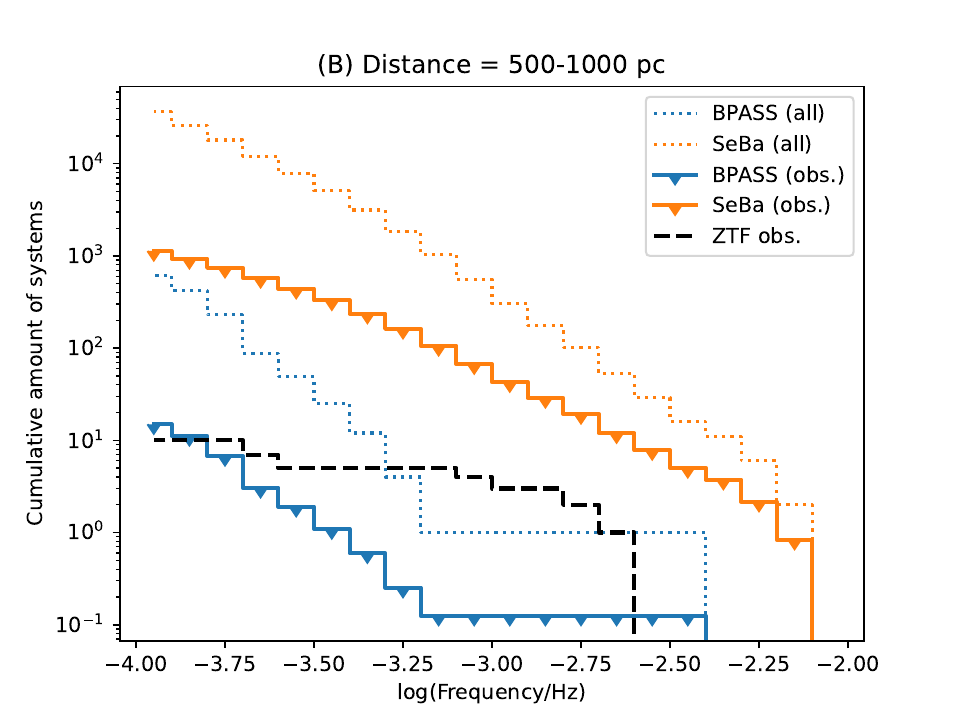}
    \includegraphics[width=0.33\linewidth]{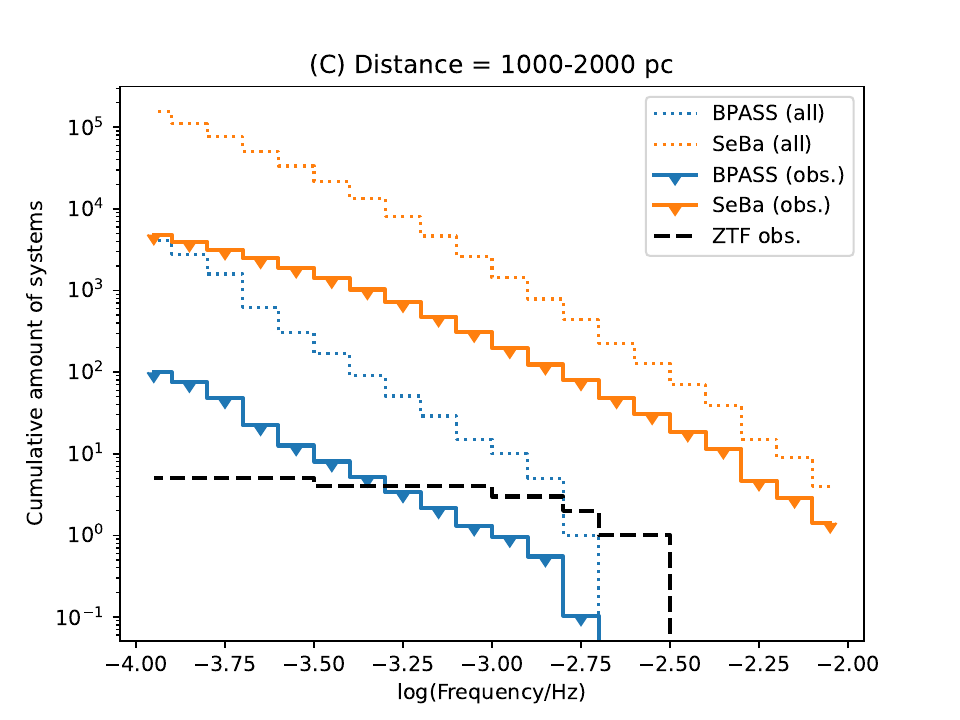}
    \caption{Cumulative sum of the amount of DWDs per frequency bin in the model populations compared to the observed population from ZTF, for several distance ranges. The cumulative sum is calculated for binaries in the frequency range of $10^{-4}$ to $10^{-2}$ Hz, starting at the high-frequency end. For the models, the dotted lines are the total numbers of DWDs while the solid lines have a correction applied to give an indication of how many DWDs would be detectable in EM when taking into account observational biases, though these are still upper limits.}
    \label{cumulative_plot}
\end{figure*}

In Fig. \ref{cumulative_plot} we compared the numbers of DWDs predicted in the two galaxy models to those in the observed ZTF dataset. Specifically, we plotted the cumulative sum of the number of DWDs per frequency bin in the frequency range of $10^{-4}$ to $10^{-2}$ Hz, starting from 0 at the high-frequency end. The different panels show the DWDs in different distance ranges - from 0 to 500 pc, from 500 pc to 1000 pc and from 1000 to 2000 pc, respectively. For the model populations, the dotted lines show the total numbers of binaries present in the galaxy model, while for the solid lines we applied a correction by weighting each system by its probability of being an eclipsing binary, using Eq. \ref{eclipse_eq} as described in Sect. \ref{method_eclipse}, and multiplying the sum of each frequency bin by a factor related to ZTF's sky coverage, as described in Sect. \ref{method_coverage}. This gives an upper limit to how many binaries in these models would actually be observed when taking into account that not all DWDs would be eclipsing and ZTF does not observe the whole sky. For the observational data, we only plotted the ZTF dataset here because, we stress again, the Munday dataset contains DWDs observed by multiple telescopes using various detection methods, and so the observational biases would be different and we could not compare the Munday data to the model populations specifically corrected for ZTF's biases.

We can see that the corrected \textsc{bpass} line is below the line of the observed ZTF binaries at high frequencies, despite this corrected line still being an upper limit for how many systems in the \textsc{bpass} model population would actually be detectable by ZTF. This trend is found in all of the distance bins, though the \textsc{bpass} line is closer to the observed line for the farthest distance bin, which is to be expected as at greater distance other observational limits such as the telescope magnitude/luminosity limit become more relevant. This indicates that \textsc{bpass} is clearly underpredicting the amount of high-frequency DWDs when compared to EM observations.

\begin{figure*}
    \centering
    \includegraphics[width=0.33\linewidth]{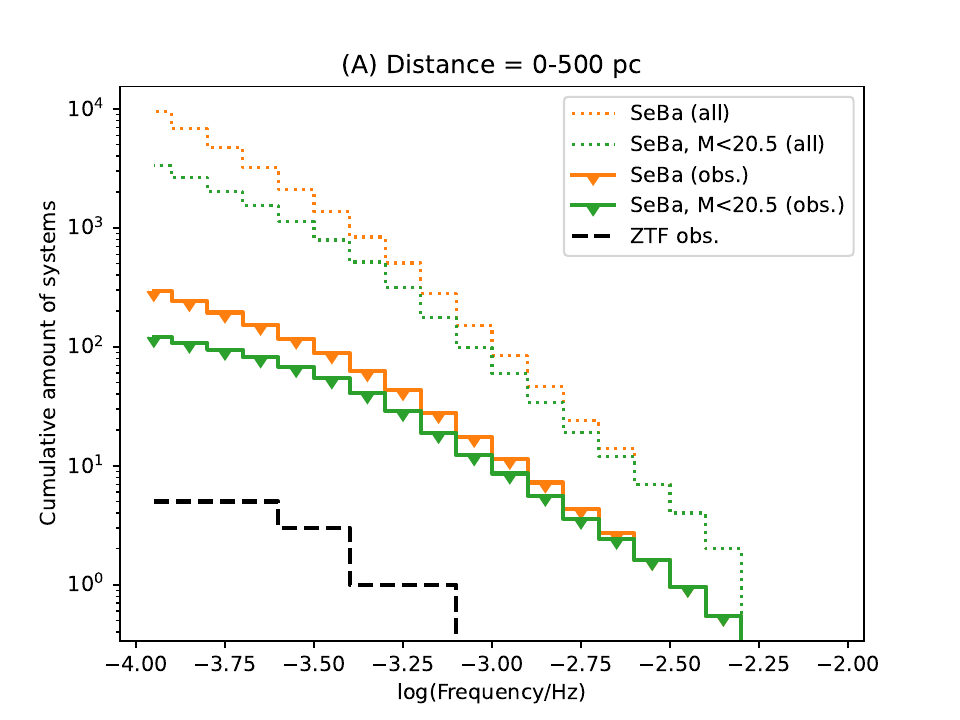}
    \includegraphics[width=0.33\linewidth]{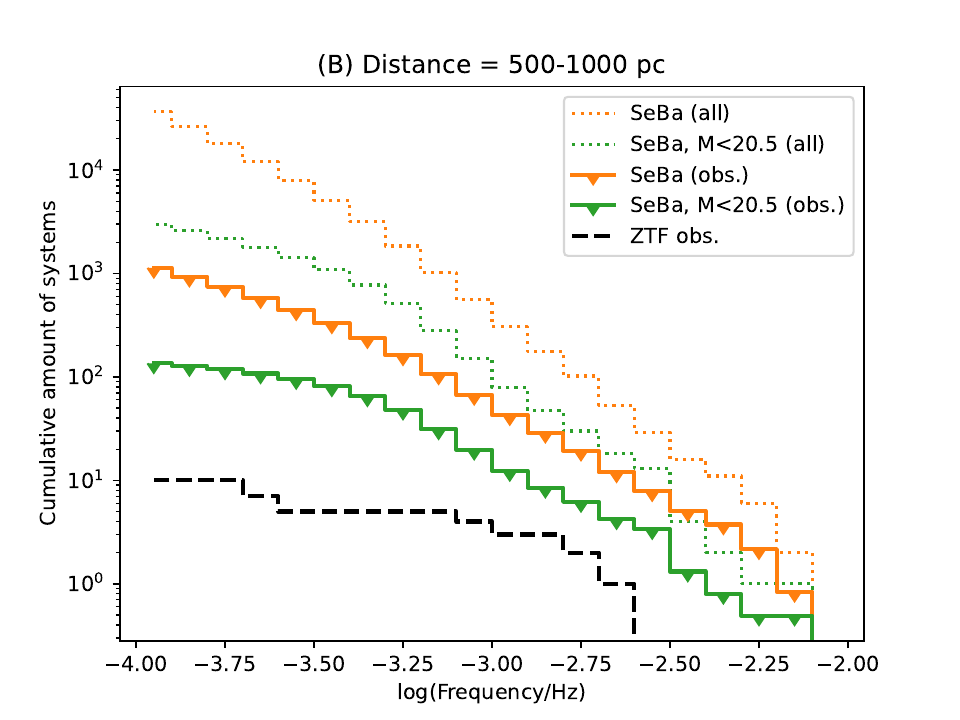}
    \includegraphics[width=0.33\linewidth]{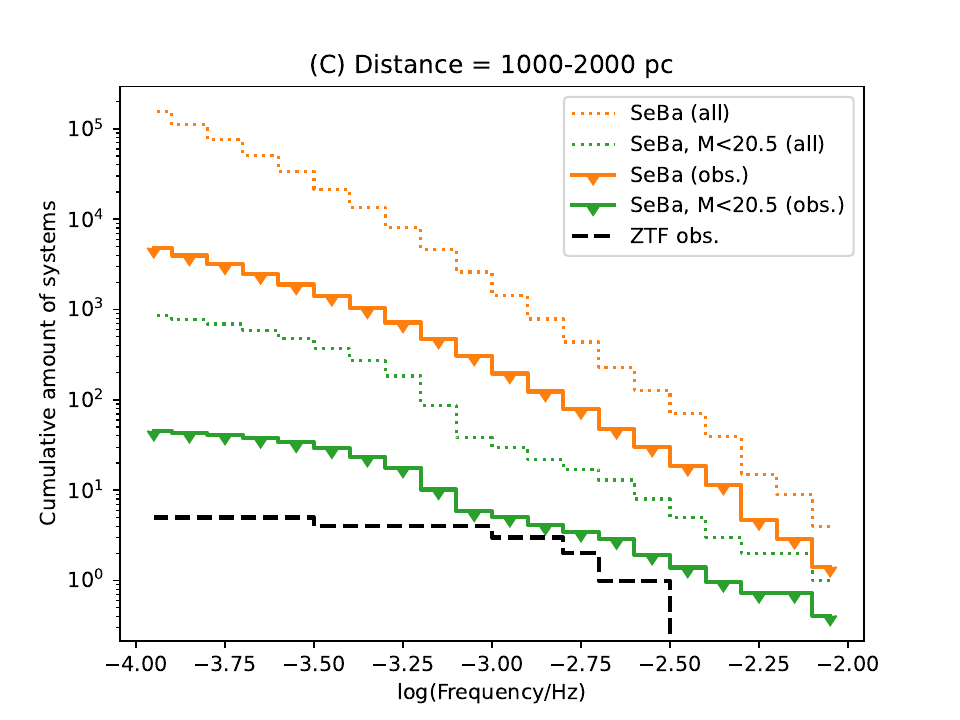}
    \caption{Comparison of the full \textsc{SeBa} galaxy model and the \textsc{SeBa} population with only DWDs brighter than magnitude 20.5 (\textit{r}-band) to the observed population from ZTF. The cumulative sum of the amount of DWDs per frequency bins is plotted, as in Fig. \ref{cumulative_plot}.}
    \label{seba21_plot}
\end{figure*}

The corrected \textsc{SeBa} lines in Fig. \ref{cumulative_plot} lie above the ZTF line for high-frequency DWDs, but as noted before this line is still an upper limit because the only observational biases being accounted for are the probability of eclipsing and the limits of ZTF's sky coverage, and so we cannot conclude whether and to what degree \textsc{SeBa} might be overpredicting compared to the observations based on this alone. To investigate this further, we took into account another limitation of the ZTF observations in Fig. \ref{seba21_plot}: that the telescope can only see the DWDs if they have sufficient brightness, or specifically if they are brighter than the limiting magnitude of ZTF, for which we took a value of 20.5 in the \textit{r}-band as descried in Sect. \ref{method_mag_limit}.

In Fig. \ref{seba21_plot} we plotted the numbers of DWDs for the total and eclipse/sky-coverage-corrected \textsc{SeBa} as in Fig. \ref{cumulative_plot}, along with a \textsc{SeBa} population which includes only DWDs brighter than magnitude 20.5, as described in Sect. \ref{method_mag_limit}. The solid green line  includes both the eclipse and sky-coverage corrections and the magnitude limit. For the closest binaries, those within 500 pc, the magnitude limit only has a small effect on the predictions and the \textsc{SeBa} predictions remain an order of magnitude above the ZTF observations, but for DWDs between 500 and 1000 pc the remaining difference is smaller. For the more distant DWDs between 1 and 2 kpc the magnitude-limited \textsc{SeBa} model is still above the ZTF line, but within a factor of a few at high frequencies.

Therefore, we can say that between 1 and 2 kpc the observability-corrected \textsc{SeBa} model population matches the observations, given the relative simple assumptions we make in our models of the DWDs' detectability. However, within 500 pc the \textsc{SeBa} galaxy model does appear to predict more DWDs than are observed in reality.

\subsection{Comparison of DWD mass distributions}

\begin{figure}
    \centering
    \includegraphics[width=\columnwidth]{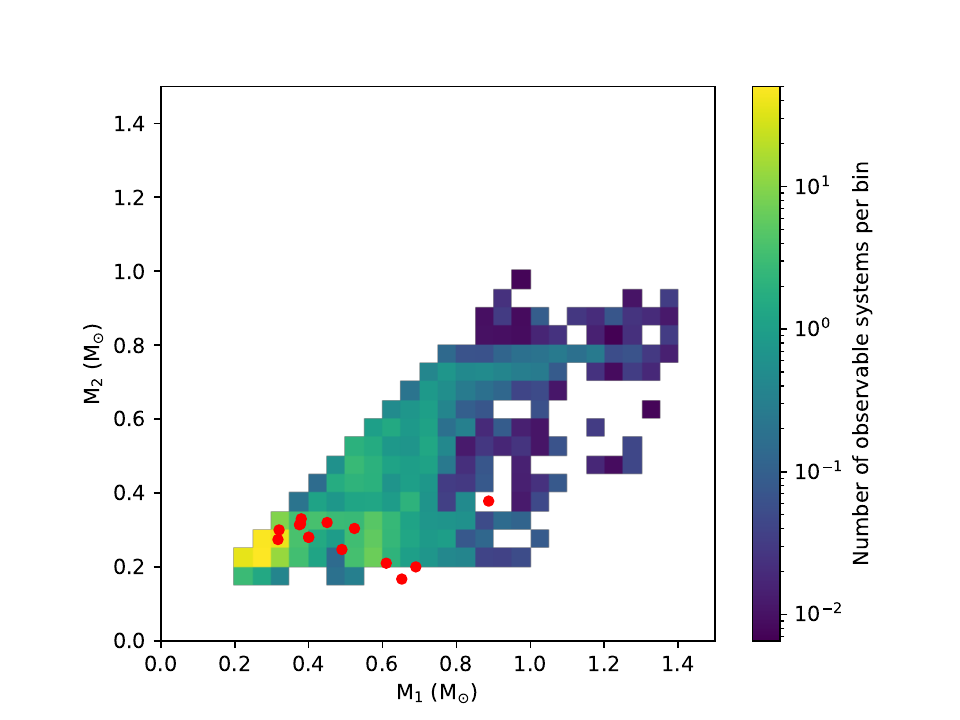}
    \caption{Comparison of the masses of the components of each DWD in the \textsc{SeBa} population with \textit{r}-band magnitude limit 20.5 to the observed population from ZTF. Each rectangular cell represents a mass bin and is shaded according to the number of systems \textsc{SeBa} predicts for that bin (corrected for the magnitude limit, probability of eclipsing and ZTF sky coverage), and each red dot is an observed system. Here, we take M$_1$ to always be the more massive of the WDs in the binary.}
    \label{mass_comp_plot}
\end{figure}

In order to check for other possible selection effects in the data (e.g. based on the masses of the WDs), we plot in Fig. \ref{mass_comp_plot} the masses of the two WDs in each binary in the \textsc{SeBa} population (with \textit{r}-band magnitude limit 20.5 and the corrections for eclipsing probability and ZTF's sky coverage applied) to those of the observed ZTF binaries. Only those DWDs in the ZTF dataset for which the masses of both components have been measured are included.

Despite the smaller ZTF sample size, we can see that the mass distributions of the predicted and observed populations resemble each other; there are no large fractions of the \textsc{SeBa} population that are unrepresented in the ZTF population or vice versa. However, the mass distributions of the \textsc{SeBa} predictions and the ZTF observations are not identical. To investigate the differences in more detail would require additional modelling of observational biases affecting the WD mass distribution, which is beyond the scope of this study.

\section{Discussion} \label{discussion_empaper}

\subsection{Implications for CEE models} \label{discussion_cee}

\citet{bpassmilkyway} and \citet{wouter_clusters_clouds} showed that the differences in the numbers of LISA-detectable DWDs between \textsc{bpass} and other stellar evolution codes like \textsc{SeBa} and \textsc{bse} are primarily caused by differences in the treatment of the stability of mass transfer and the modelling of CEEs. Firstly, mass transfer is more likely to be stable in \textsc{bpass} than in other codes \citep{bpassmasstransfer,wouter_clusters_clouds}, resulting in fewer CEEs occurring. Secondly, the method by which CEEs themselves are modelled in \textsc{bpass} is qualitatively different.

Rapid population synthesis codes such as \textsc{SeBa} and \textsc{bse} use analytic prescriptions to model the outcomes of CEEs, of which the most common are the $\alpha$-formalism \citep{webbink1984} and the $\gamma$-formalism \citep{nelemans_gamma1,nelemans_gamma2}, though there are also alternative prescriptions such as those of \citet{trani2022} and \citet{hirai2022}. In \textsc{bpass}, by contrast, the stellar structure is evolved in detail throughout the CEE \citep{bpass1,bpassgw170817}. However, the detailed CEE model of \textsc{bpass} is still dependent on assumptions and simplifications about the physics of stellar interactions, and so its results are not necessarily more accurate or realistic.

In particular, CEEs in \textsc{bpass} tend to be more efficient than in \textsc{SeBa} and \textsc{bse}, which means that the binaries tend to lose less angular momentum during the CEE and thus undergo less orbital shrinkage \citep{bpassgw170817}, resulting in wider orbits at the end of the CEE, which reduces the amount of DWDs emitting GWs in the LISA frequency band compared to other codes \citep{bpassmilkyway,wouter_clusters_clouds}. When compared to $\alpha$ and $\gamma$-prescriptions, the effective $\alpha \lambda$ and $\gamma$ parameters of the \textsc{bpass} CEEs are much higher than those used in \textsc{SeBa}; for example, in \textsc{bpass} the effective $\alpha \lambda$ value tends to be between 8 and 20 for CEE in the first binary interaction and between 4 and 30 for the second \citep{wouter_clusters_clouds}, while the \textsc{SeBa} version of \citet{seba_toonen} assumes $\alpha \lambda$ = 2, which means that the \textsc{bpass} CEE model is an order of magnitude more efficient.

The \textsc{SeBa} values for $\alpha \lambda$ and $\gamma$ were calculated based on fits to astrophysical observations of DWDs \citep{nelemans_gamma1,nelemans_seba}, which raises the question of whether the much higher values from the \textsc{bpass} CEE model are not representative in this case. Our results in this study, in which \textsc{bpass} clearly underpredicts short-period DWDs compared to observations, suggest that the \textsc{bpass} CEE model is indeed too efficient. However, we note that in this study we are only testing the CEE models for low- and intermediate-mass stars, as massive stars do not produce WDs. Furthermore, \citet{bpassmasstransfer} found that \textsc{bpass}'s predictions for massive binaries do match the rates of BH binaries detected by LVK. Therefore, it is important to consider that modelling stars in different mass ranges may require different CEE prescriptions.

Additionally, these results show that LISA, which will identify a far larger number of Galactic DWDs than the EM surveys we used \citep{lisa_astrophysics}, will provide invaluable information on the processes of mass transfer and CEEs. This aligns with the results of \citet{delfavero2025}, who found that LISA's future observations of the Galactic DWD population will be able to constrain CEE efficiency parameters to a level of about 10\%.

\subsection{Additional observational limiting factors}

Our estimates of how many DWDs in the model populations would be detectable by ZTF through eclipses are upper limits, as there are additional factors which limit the detections of the DWDs and their eclipses.

One of these relates to eclipse depth, which measures how much a DWD's apparent brightness decreases during an eclipse. Eq. \ref{eclipse_eq} gives the probability of an eclipse of any depth, but ZTF could only identify an eclipse as such if the eclipse depth is at least approx. 10\% of the total brightness \citep{elbadry2022}. The eclipse depth is not trivial to calculate, as it depends on the luminosities and temperatures of the WDs, which in turn depend on modelling of WD cooling.

The implication of the eclipse depth threshold is that the amount of DWDs in a population that ZTF would actually be able to identify as eclipsing binaries would be slightly lower than the total numbers of eclipsing binaries as plotted in Figs. \ref{cumulative_plot} and \ref{seba21_plot}.

Another limiting factor is that DWDs are not distributed isotropically across the sky, but are most likely to be found in the MW disk \citep[e.g.][]{lamberts2019,breivik2020}, which is where the sky density of stars is also highest. This high density results in crowding, where the close proximity of multiple different light sources makes it more difficult to resolve the light of each individual one \citep[e.g.][]{scheuer1957,condon1974,renzini1998,olsen2003}.

\begin{figure*}
    \centering
    \includegraphics[width=0.33\linewidth]{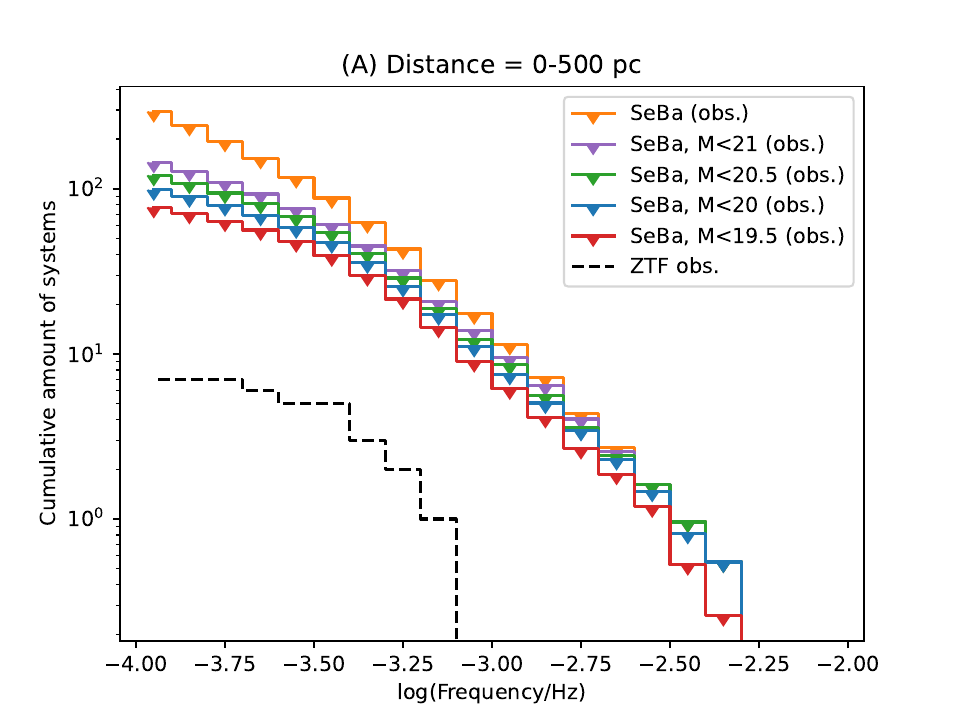}
    \includegraphics[width=0.33\linewidth]{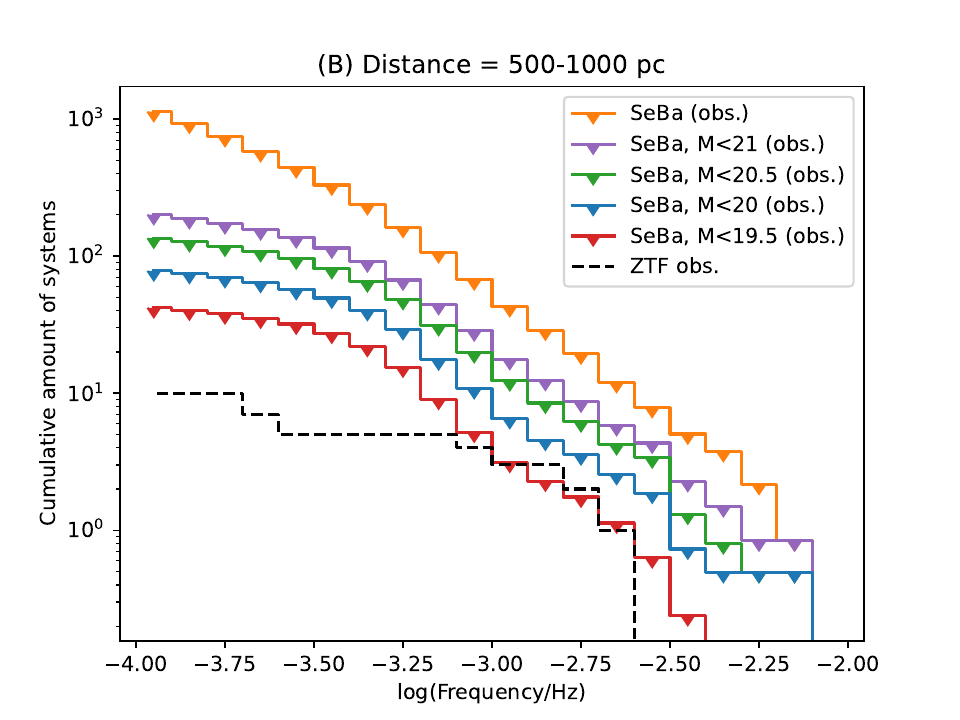}
    \includegraphics[width=0.33\linewidth]{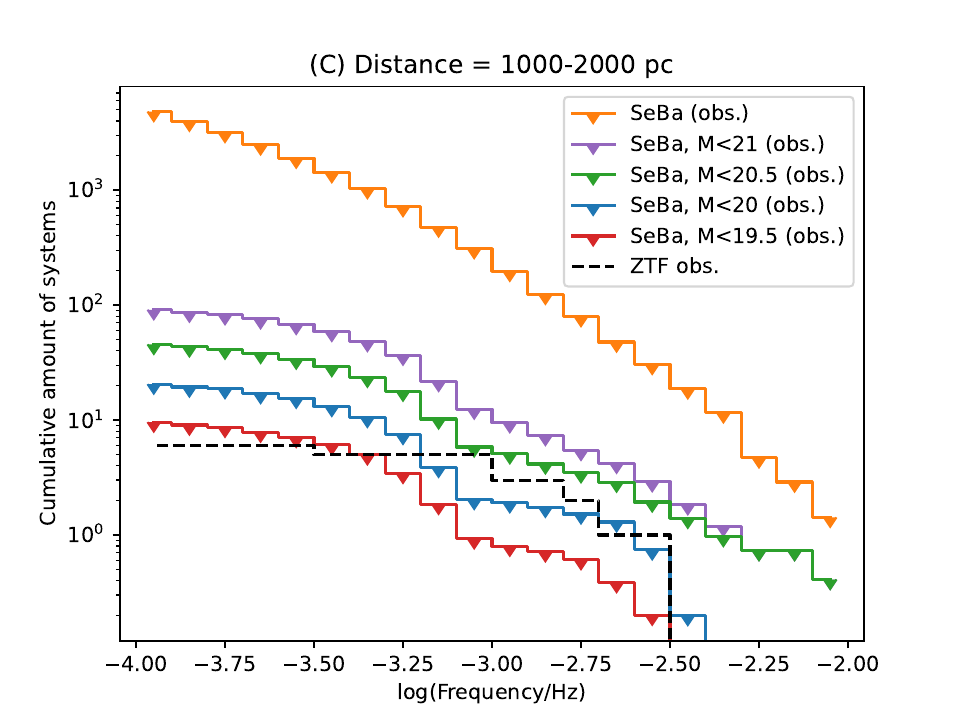}
    \caption{Comparison of the \textsc{SeBa} galaxy model with only DWDs brighter than various (\textit{r}-band) magnitude limits to the observed population from ZTF. The cumulative sum of the amount of DWDs per frequency bins is plotted, as in Fig. \ref{cumulative_plot}. All lines are corrected for eclipsing probability and ZTF's sky coverage.}
    \label{seba_mag_plot}
\end{figure*}

As a result, the threshold magnitude at which ZTF could detect DWDs would effectively be somewhat brighter than previously assumed for sources in the Galactic plane, which would lower the numbers of DWDs in the model populations that are bright enough to be detected by ZTF. In Fig. \ref{seba_mag_plot} we plotted what the magnitude-limited \textsc{SeBa} population from Sect. \ref{method_mag_limit} and Fig. \ref{seba21_plot} looks like when we vary the magnitude limits between 21 and 19.5. We can see that, for binaries closer than 500 pc, the \textsc{SeBa} predictions remain an order of magnitude above the ZTF observations, with a magnitude limit of 19.5 producing about a factor of two fewer detectable DWDs than a limit of 20.5. However, for binaries between 500 and 1000 pc, the \textsc{SeBa} predictions match or even fall below the ZTF observations if the magnitude limit is 19.5 or brighter, at least at high frequencies, and for binaries between 1000 to 2000 pc this is the case for a magnitude limit of 20 or brighter.

While we use a default magnitude limit of 20.5, a limit of 20 is not unreasonable in the presence of crowding and extinction in the MW disk, where the largest concentration of DWDs reside. Therefore, the \textsc{SeBa} predictions are consistent with the ZTF observations of DWDs beyond 500 pc, but the \textsc{SeBa} model does appear to overpredict the number of observable DWDs within 500 pc by up to an order of magnitude. Because of the distance dependence, it is possible that the overprediction at low distances is caused by the spatial distribution of the MW model used by \citet{korol2017}, as opposed to the intrinsic stellar evolution modelling of \textsc{SeBa}.

Specifically, \citet{korol2017} assumed a SFH from \citet{boissier1999} and an axisymmetric disk structure as detailed in \citet{nelemans2004}, where the stellar density is homogeneous for all points within the disk at a given distance from the Galactic centre, and so features such as spiral arms are not taken into account, which could have an effect on the density of DWDs close to the Earth. In particular, part of the Perseus Arm is within 2 kpc of Earth \citep{xu2006}, which could contribute to the number of DWDs in the distance bin between 1 and 2 kpc but not to the more nearby bins.

However, \citet{bpassmilkyway} used a non-axisymmetric model for the structure of the MW disk, and they found that if they varied the location of the Earth between different points at the same Galactic radius, the changes in the WD density were small, with the amount of LISA-detectable DWDs varying by only a few percent. Therefore, it is not clear whether \citet{korol2017}'s assumption of axisymmetry would have a large effect on the local density of DWDs. There are also other differences between the MW models used by \citet{korol2017} and \citet{bpassmilkyway} that could affect the local stellar density, such as the disk scale parameters which determine the thickness and distribution of the disk, but a detailed examination of these MW structure models is beyond the scope of this paper.

\subsection{Uncertainties in population synthesis predictions}

There are some uncertainties in the process by which we translate the \textsc{bpass} and \textsc{SeBa} galaxy models into predictions of EM-detectable DWDs. Firstly, the WD radii we calculate in Sect. \ref{method_eclipse} are approximate and based on a fit to the manual radii in the ZTF dataset; as can be seen in Fig. \ref{mass_radius_plot}, there are a few ZTF WDs that are outliers with respect to the mass-radius relationship, and there generally is a degree of scatter in the ZTF measurements which creates uncertainty in the mass-radius relationship. This in turn creates uncertainty in the eclipse probabilities for each of the model systems.

Secondly, as noted in Sect. \ref{method_mag_limit}, the WD magnitudes from \textsc{SeBa} were computed using cooling models that do not necessarily have radii that are consistent with what we used to calculate the probability of a DWD being eclipsing. Given the counterbalancing effects of larger radii leading to brighter WDs but also to faster cooling, we believe that the effect of the larger radii is (much) less important for the magnitude cuts than for the eclipse probability.

Another source of uncertainty is the cooling models themselves, as these assume that all WDs have a carbon–oxygen (CO) chemical composition \citep{toonen2013}. Thus, for WDs with masses below approx. 0.5 M$_{\odot}$ or above 1.1 M$_{\odot}$, which would be expected to have different chemical compositions in reality, the calculated luminosities are less accurate.

\subsection{Using the Munday data set as a magnitude-limited sample}

When we compared the frequency distributions of the DWDs in the population synthesis models to those in the EM observations, we only used the systems in the ZTF data set because, as previously mentioned, the Munday data set contains observations from multiple different telescopes, and from both eclipsing DWDs and DWDs detected through radial velocity measurements. The latter category makes up a majority of the Munday DWDs, particularly at frequencies lower than 10$^{-3}$ Hz.

DWDs detected through radial velocity measurements usually have \textit{g}-band magnitudes of 18 or brighter \citep[see e.g.][]{munday2024}. One may wonder if it is possible to use the Munday data set as a complete magnitude-limited sample in a comparison with the population synthesis models. However, the Munday data set contains only 41 DWDs with a \textit{g}-band magnitude of 18 or brighter, a GW frequency of 10$^{-4}$ Hz or higher and a distance of no more than 1 kpc from Earth. By comparison, the \textsc{SeBa} galaxy model contains 791 such DWDs. Similarly, using a \textit{g}-band magnitude limit of 16, the Munday data set has 15 DWDs while the \textsc{SeBa} galaxy model has 96.

This indicates that the Munday data set is quite incomplete in terms of capturing the full population of DWDs brighter than a certain threshold. Therefore, while the Munday data set is highly interesting in terms of learning about properties of DWDs, we cannot use it for evaluating the total number or space density of DWDs, which is the aim of this paper.

\section{Conclusions} \label{conclusions_empaper}

We have taken the DWDs predicted by the \textsc{bpass} galaxy model of \citet{bpassmilkyway} and the \textsc{SeBa} galaxy model of \citet{korol2017}, and compared these to the DWDs observed by EM surveys. Compared to the observations, we have found that the \textsc{bpass} model underpredicts short-period (high-frequency) DWDs by a factor of at least ten. By contrast, the \textsc{SeBa} model is roughly consistent with the DWD observations for systems between 500 pc and 1 kpc and clearly consistent for those beyond 1 kpc, but may overpredict the number of DWDs within 500 pc, though by no more than a factor of ten; due to the uncertainties in the selection effects we cannot make a more definitive statement about the \textsc{SeBa} model.

These results imply that the total number of resolved LISA DWDs will likely be on the order of 10,000, as opposed to less than 1000 as predicted by \citet{bpassmilkyway}. There may be additional selection effects that we have not modelled, which would decrease the DWD rates from the population synthesis models compared to the observations, which would increase the factor by which the \textsc{bpass} model is underpredicting and decrease the factor by which the \textsc{SeBa} model may be overpredicting.

As the treatment of mass transfer and CEEs in \textsc{bpass} compared to \textsc{SeBa} is the main factor causing the difference in the predictions of short-period, LISA-detectable DWDs between these codes \citep{wouter_clusters_clouds}, this study forms a valuable test of common-envelope prescriptions, showing that the \textsc{bpass} CEE prescription is too efficient for low- and intermediate-mass stars. As earlier research by \citet{bpassmasstransfer} showed that \textsc{bpass} predictions are consistent with GW observations for massive (BH-forming) stars, this indicates that modelling stars in different mass ranges may require different CEE prescriptions. Additionally, our findings highlight how LISA's observations of DWDs will provide invaluable information on mass transfer and CEEs.

\section*{Data availability}

The \textsc{bpass} models can be downloaded from \url{https://bpass.auckland.ac.nz}. The \citet{munday2024} observed DWD catalogue can be downloaded from \url{https://github.com/JamesMunday98/CloseDWDbinaries}. Other data underlying this article will be shared upon reasonable request to the corresponding author.

\begin{acknowledgements}

The authors thank the anonymous referee for their comments, and Paul J. Groot, James Munday and Simon F. Portegies Zwart for useful discussions. WGJvZ acknowledges support from Radboud University, the European Research Council (ERC) under the European Union’s Horizon 2020 research and innovation programme (grant agreement No.~725246), Dutch Research Council grant 639.043.514 and the University of Auckland. JJE acknowledges support from Marsden Fund Council grant MFP-UOA2131 managed through the Royal Society of New Zealand Te Apārangi.

\end{acknowledgements}

\bibliographystyle{aa}
\bibliography{references}

\end{document}